\documentclass[notitlepage,showpacs]{revtex4-1}
\usepackage[english]{babel}
\usepackage{amsmath}
\usepackage{amssymb}
\usepackage{bm}
\usepackage{graphicx}
\usepackage{color}
\usepackage{float}
\usepackage{siunitx}
\usepackage[capitalize]{cleveref}
\usepackage[caption=false]{subfig}

\DeclareMathOperator{\D}{\overset{\leftrightarrow}{D}}
\DeclareMathOperator{\cc}{c.c.}

\begin{document}

\title{Nonlinear wave damping due to multi-plasmon resonances}
\author{G.Brodin, R. Ekman, J. Zamanian}
\affiliation{Department of Physics, Ume{\aa } University, SE--901 87}

\begin{abstract}
For short wavelengths, it is well known that the linearized Wigner-Moyal
equation predicts wave damping due to wave-particle interaction, where the
resonant velocity shifted from the phase velocity by a velocity $v_q=\hbar
k/2m$. Here $\hbar$ is the reduced Planck constant, $k$ is the wavenumber
and $m$ is the electron mass. 
Going beyond linear theory, we find additional resonances with velocity
shifts $nv_q$, $n = 2, 3, \ldots$, giving rise to a new wave-damping
mechanism that we term \emph{multi-plasmon damping}, as it can be seen as
the simultaneous absorption (or emission) of multiple plasmon quanta.
Naturally this wave damping is not present in classical plasmas. For a
temperature well below the Fermi temperature, if the linear ($n = 1$)
resonant velocity is outside the Fermi sphere, the number of linearly
resonant particles is exponentially small, while the multi-plasmon
resonances can be located in the bulk of the distribution. We derive sets of
evolution equations for the case of two-plasmon and three-plasmon resonances
for Langmuir waves in the simplest case of a fully degenerate plasma. By
solving these equations numerically for a range of wave-numbers we find the
corresponding damping rates, and we compare them to results from linear
theory to estimate the applicability. Finally, we discuss the effects due to
a finite temperature.
\end{abstract}

\maketitle

\affiliation{Department of Physics, Ume{\aa } University, SE--901 87 Ume{\aa
}, Sweden,}


\section{Introduction}

Wave-particle interaction has been of long-standing interest in plasma
physics, where Landau damping of Langmuir waves has been a prominent
example. In a classical context Landau damping has been studied extensively,
both linearly as well as in the nonlinear regime when effects such as bounce
oscillations enters the picture, see e.g. Refs.~\cite{Nicholsson,1977,
Brodin-1997,Manfredi-1997,Danielson2004,PRL-2009}. 
Including quantum effects, new mechanisms come into play, for example the
effect of the electron spin \cite{Asenjo,Zamanian 2010}, nonlinear
modifications of the Fermi surface \cite{Brodin-Stefan}, quantum
modifications of the radiation pressure \cite{Stefan-2011}, exchange effects
~\cite{Andreev-2014, Ekman 2015} and bounce-like oscillations in the absence
of trapped particles~\cite{Scripta-2015}, even in a \textquotedblleft
weak\textquotedblright\ quantum regime. It can also be noted that the
classical-quantum transition of plasmons recently has been probed
experimentally, using metallic nanoparticles \cite{Nanoparticles}.

In the present paper we will instead be interested in a ``strong'' quantum
regime~\cite{Haas-2001,Eliasson-Shukla-JPP,Rightley-Uzdensky}. By weak
(strong) we mean that the Langmuir wavelength is much less than (comparable
to) the characteristic de Broglie wavelength. 
The main difference between the weak and strong quantum regimes is that in
the former, the resonant particles still have a velocity close to the phase
velocity of the wave, just as in classical theory. Nevertheless it should be
stressed that significant effects, e.g. quantum suppression of nonlinear
bounce oscillations~\cite{Scripta-2015, Daligault, Feix-1991}, may occur
already in the weak quantum regime. In the strong quantum regime the
resonant particles instead fulfill $\omega - kv_{z} \pm \hbar k^{2}/2m
\simeq 0$ in linearized theory~\cite{Eliasson-Shukla-JPP, Rightley-Uzdensky,
Misra-preprint, Mendonca-2016}. Here $\omega $ and $\mathbf{k} = k\mathbf{%
\hat{z}}$ are the wave frequency and wavevector, $v_{z}$ is the electron
velocity along $\hat{z}$, $h = 2\pi \hbar $ is Planck's constant, and $m$ is
the electron mass.

We will demonstrate that if we study the nonlinear evolution, there appear 
\emph{multi-plasmon resonances}, that is, resonant wave-particle interaction
with a resonant velocity 
\begin{equation}
v_{\text{res},\pm n}=\frac{\omega }{k}\pm \frac{n\hbar k}{2m},
\label{eq:resonancecondition}
\end{equation}%
where $n$ is a positive integer. The case with $n=1$ corresponds to
one-plasmon processes (linear Landau damping), $n=2$ corresponds to
two-plasmon processes, etc. 
The physical meaning of the resonance condition~\eqref{eq:resonancecondition}
and the integer $n$ can be understood as follows. When a particle absorbs or
emits a wave quantum its momentum can increase or decrease according to 
\begin{equation}
\hbar k_{1}\pm \hbar k=\hbar k_{2}  \label{eq:momentumTransfer}
\end{equation}%
and at the same time the energy changes according to 
\begin{equation}
\hbar \omega _{1}\pm \hbar \omega =\hbar \omega _{2}.
\label{eq:energyTransfer}
\end{equation}%
Next we identify $\hbar k_{1}/m$ (or equally well $\hbar k_{2}/m$) with the
resonant velocity, and note that for small amplitude waves the particle
frequencies and wavenumbers $(\omega _{1,2},k_{1,2})$ obey the free particle
dispersion relation $\omega _{1,2}=\hbar k_{1,2}^{2}/2m$, that we get
directly from the single particle Schr\"{o}dinger equation by letting the
wave field go to zero. Using these relations we see that the energy momentum
relations (\ref{eq:momentumTransfer}) and (\ref{eq:energyTransfer}) imply
the quantum modification of the resonant velocity as seen in Eq. (\ref%
{eq:dispRel-2}), from linearized theory. The linear quantum modifications of
the resonant denominators is what is obtained by putting $n=1$ in Eq.~%
\eqref{eq:resonancecondition}.

A possibility that requires nonlinear theory is the simultaneous absorption
of multiple wave quanta, rather than a single wave quanta at a time. In that
case \cref{eq:momentumTransfer,eq:energyTransfer} are replaced by 
\begin{equation}
\hbar k_{1}\pm n\hbar k=\hbar k_{2}  \label{eq:momTransferMulti}
\end{equation}
and 
\begin{equation}
\hbar \omega _{1}\pm n\hbar \omega =\hbar \omega _{2}
\label{eq:energyTransferMulti}
\end{equation}%
where $n=1,2,3...$ is an integer. Using %
\cref{eq:momTransferMulti,eq:energyTransferMulti} together with the free
particle dispersion relations we recover~\cref{eq:resonancecondition}). Thus
it is clear that the integer $n$ in this equation represents the number of
wave quanta that is simultaneously absorbed or emitted. When we pick the
minus sign in (\ref{eq:resonancecondition}) the resonant velocity for
absorbing multiple wave quanta can be considerably smaller, provided the
wavelengths are short. 

For pedagogical reasons, and as the $n = 1$ resonance is well-known already
from linear theory, we will treat a fully degenerate plasma and assume that
the wavenumber $k$ is such that one-plasmon processes are forbidden due to a
lack of resonant particles. This is the case if $k < k_\text{cr}$ where $k_%
\text{cr}$ is the critical wavenumber computed in Ref.~\cite%
{Eliasson-Shukla-JPP}. 
Specifically we will focus on two-plasmon and three-plasmon processes, that
can be studied if we include up to cubic terms in an amplitude expansion.
The damping rates of Langmuir waves, which decrease with the decaying
amplitude, are found both for $n = 2$ and $n = 3$, and the evolution of the
Wigner function is computed numerically. We stress here that wave damping
due to multi-plasmon resonances (i.e. $n\geq 2$) is a fundamentally new
process. Contrary to ordinary Landau damping it does not exist in the limit $%
\hbar \to 0$. However, as indicated by the above discussion, the new
resonances are present also for completely non-degenerate systems. The
effects of a finite temperature will be discussed in the final section. 


\section{Basic equations and preliminaries}

We take as our starting point the Wigner-Moyal equation for electrons 
\begin{equation}
\frac{\partial f}{\partial t} + \mathbf{v }\cdot \nabla_{\mathbf{r}} f - 
\frac{iq m^3}{\hbar} \int \frac{d^3\mathbf{r^{\prime }} d^3\mathbf{v^{\prime
}}}{(2\pi\hbar)^3} e^{i\mathbf{r^{\prime }}\cdot(\mathbf{v }- \mathbf{%
v^{\prime }})m/\hbar} \big[ \Phi(\mathbf{r }+ \frac{\mathbf{r^{\prime }}}{2}%
)- \Phi(\mathbf{r }- \frac{\mathbf{r^{\prime }}}{2}) \big ] f(\mathbf{r},%
\mathbf{v^{\prime }},t) = 0  \label{eq:wigner}
\end{equation}

Here $f$ is the Wigner function, $q = -|e|$ is the electron charge, and we
have restricted ourselves to the electrostatic case, where $\Phi$ is the
scalar potential. The system is closed with Poisson's equation, 
\begin{equation}
- \nabla^2 \Phi = \frac{q}{\epsilon_0} \int d^3\mathbf{v} f - \frac{q}{%
\epsilon_0} n_{i}  \label{eq:poisson}
\end{equation}
where $n_{i}$ is a constant neutralizing ion background.%

\subsection{Short review of linearized theory}

While the theory of multi-plasmon resonances requires a study of the \textit{%
nonlinear} Wigner equation, there is much information that follows directly
from the linearized theory. In particular the parameters encountered in the
nonlinear theory are functions of $\omega $ and $k$, whose relation is
determined by the linear dispersion relation to leading order in an
amplitude expansion. A highly useful study of linear theory is provided by
Ref. \cite{Eliasson-Shukla-JPP}. Our presentation here will briefly review
some of these results, but also point out a few aspects of the linear theory
that are of special relevance for the (nonlinear) multi-plasmon processes.

Dividing the Wigner function as $f=F_{0}+f_{1}(\mathbf{v})\exp [i(kz-\omega
t)]$ where $F_{0}$ is the background distribution and linearizing %
\cref{eq:wigner}, the solution for the Wigner function is%
\begin{equation}
f_{1}=-\frac{q\Phi \left[ F_{0}(\mathbf{v}+\mathbf{v}_{q})-F_{0}(\mathbf{v}-%
\mathbf{v}_{q})\right] }{\hbar (\omega -kv_{z})},  \label{lin-1}
\end{equation}%
where we have introduced $\mathbf{v}_{q}=v_{q}\mathbf{\hat{z}}=(\hbar k/2m)%
\mathbf{\hat{z}}$, see Ref. \cite{Eliasson-Shukla-JPP} for further details.
Inserting this expression into Poisson's equation (\ref{eq:poisson}), the
linear dispersion relation becomes%
\begin{equation}
1=-\frac{q^{2}}{\hbar k^{2}\varepsilon _{0}}\int \frac{F_{0}(\mathbf{v}+%
\mathbf{v}_{q})-F_{0}(\mathbf{v}-\mathbf{v}_{q})}{(\omega -kv_{z})}\,d^{3}v.
\label{eq:dispRel-1}
\end{equation}%
Changing integration variables the dispersion relation can be written as 
\begin{equation}
1+\frac{q^{2}}{\hbar k^{2}\varepsilon _{0}}\int \left( \frac{F_{0}(\mathbf{v}%
)}{\omega -kv_{z}+\hbar k^{2}/2m}-\frac{F_{0}(\mathbf{v})}{\omega
-kv_{z}-\hbar k^{2}/2m}\right) \,d^{3}v=0.  \label{eq:dispRel-2}
\end{equation}%
Taking $F_{0}$ as a Fermi-Dirac distribution with $T=0$ we have $F_{0}=0$
for $\left\vert \mathbf{v}\right\vert >v_{F}$ and $F_{0}=2m^{3}/(2\pi \hbar
)^{3}$for $\left\vert \mathbf{v}\right\vert \leq v_{F}$, where $%
v_{F}=(2k_{B}T_{F}/m)^{1/2}$ is the Fermi velocity. In accordance with the
assumption of no linear poles, we will limit ourselves to values of $k$ such
that $F_{0}=0$ whenever the denominator in the integral is zero. Some
intermediate steps to get the solutions for the integrals are given in Ref. 
\cite{Eliasson-Shukla-JPP}. Here we just write out the final form for the
dispersion relation that applies in the absence of linear poles. The
dispersion relation is then given as 
\begin{equation}
1+\chi _{e}=0,  \label{eq:dispRel-3}
\end{equation}%
where the linear susceptibility $\chi _{e}$ is given by%
\begin{equation*}
\begin{aligned} \chi _{e} &=&\frac{3\omega
_{pe}^{2}}{4k^{2}v_{F}^{2}}\left\{ 2-\frac{m_{e}}{\hbar kv_{F}}\left[
v_{F}^{2}-\left( \frac{\omega }{k}+\frac{\hbar k}{2m}\right) ^{2}\right] \ln
\left\vert \frac{\frac{\omega }{k}-v_{F}+\frac{\hbar k}{2m}}{\frac{\omega
}{k}+v_{F}+\frac{\hbar k}{2m}}\right\vert \right. \\ &&+\left.
\frac{m_{e}}{\hbar kv_{F}}\left[ v_{F}^{2}-\left( \frac{\omega
}{k}-\frac{\hbar k}{2m}\right) ^{2}\right] \ln \left\vert \frac{\frac{\omega
}{k}-v_{F}-\frac{\hbar k}{2m}}{\frac{\omega }{k}+v_{F}-\frac{\hbar
k}{2m}}\right\vert \right\} . \end{aligned}
\end{equation*}

\begin{figure}[tbp]
\includegraphics[width=\linewidth]{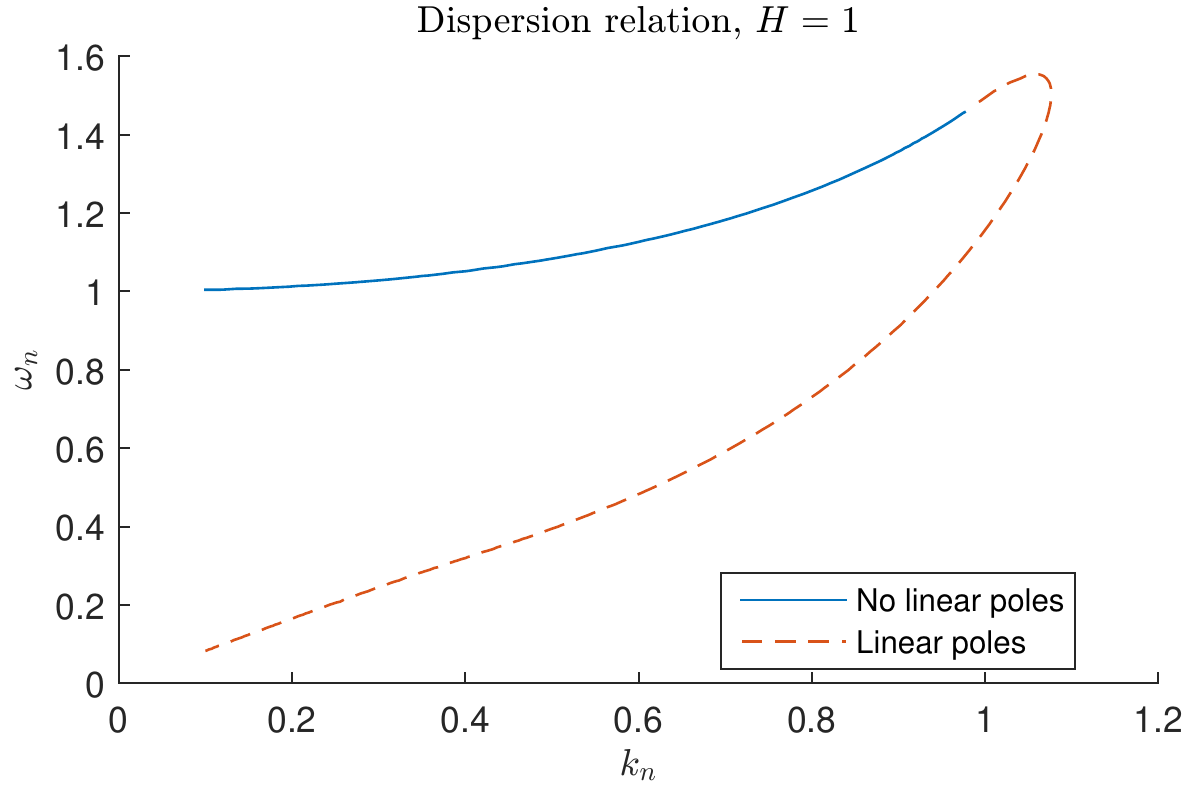}
\caption{A numerical solution of the dispersion relation for $H=1$. The
dashed part (red online) has pole contributions, which are not included in
the expression for the susceptibility. Hence only the solid part of the
curve (blue online) is formally valid. The first part of the dashed curve
that connects to the solid line has a small pole contribution. Hence this
part still gives an approximately correct expression for the real part of $%
\protect\omega$. }
\label{fig:dispersionRelation}
\end{figure}

\subsection{Numerical solutions - location of multi-plasmon resonances}

Now we want to separate the wavenumber spectrum into different regimes,
depending on what type of resonances that occur. For this purpose we focus
on the negative sign in~\cref{eq:resonancecondition}, as this gives the
lowest resonant velocity. Moreover, we introduce the normalized frequency $%
\tilde{\omega}=\omega /\omega _{p}$, and the normalized wavenumber $\tilde{k}%
=\hbar k/mv_{F}$, in which case $\chi _{e}$ obtains the form 
\begin{equation}
\begin{aligned} \chi _{e} =&\frac{3H^{2}}{4\tilde{k}^{2}}\left\{
2-\frac{1}{\tilde{k}}\left[ 1-\left(
H\frac{\tilde{\omega}}{\tilde{k}}+\frac{\tilde{k}}{2}\right) ^{2}\right] \ln
\left\vert
\frac{H\frac{\tilde{\omega}}{\tilde{k}}-1+\tilde{k}/2}{H\frac{\tilde{%
\omega}}{\tilde{k}}+1+\tilde{k}/2}\right\vert \right. \\ & + \left.
\frac{1}{\tilde{k}}\left[ 1-\left(
H\frac{\tilde{\omega}}{\tilde{k}}-\frac{\tilde{k}}{2}\right) ^{2}\right] \ln
\left\vert
\frac{H\frac{\tilde{\omega}}{\tilde{k}}-1-\tilde{k}/2}{H\frac{\tilde{%
\omega}}{\tilde{k}}+1-\tilde{k}/2}\right\vert \right\} , \end{aligned}
\label{suscept-1}
\end{equation}%
where $H=\hbar \omega _{p}/mv_{F}^{2}$. Since $H$ scales with density only
as $n^{-1/6}$, and is of order unity for a wide range of densities including
metallic densities, we will here limit ourselves to the case $H=1$. A few
numerical tests in the range $\,0.5\leq H\leq 2$ have been made, and the
conclusions remain qualitatively the same. \ A numerical solution $\omega
=\omega (k)$ for $H=1$ is given in \cref{fig:dispersionRelation}. The
solution only applies provided the condition of no linear poles is met, and
the part which violates this condition is indicated. Still it should be
noted that the dashed part of the curve gives a good description of the
electron-acoustic branch of the dispersion relation provided that
wave-particle interaction is not too strong (see e.g. Ref. \cite%
{electron-acoustic} for a discussion of the electron-acoustic branch). While
the electron-acoustic branch also is subject to multi-plasmon resonances, we
will be concerned with the Langmuir branch from now on. Our purpose is to
avoid the presence of linear resonances altogether in order to focus our
attention on physics induced by the multi-plasmon resonances. Since we are
solely focused on the case without linear poles, this means that we assume $%
v_{\text{res,-1}}>v_{F}$, where we apply the negative sign and $n=1$ in %
\cref{eq:resonancecondition}. Assuming that both $\omega $ and $k$ are
positive and keeping $H=1$ the inequality $v_{\text{res,-1}}>v_{F}$ implies 
\begin{equation}
\tilde{\omega}-\tilde{k}-\tilde{k}^{2}/2>0.  \label{Cond-1}
\end{equation}%
First focusing on the case of two-plasmon resonance we want to combine (\ref%
{Cond-1}) with the presence of two-plasmon resonances, that is $v_{\text{%
res,-2}}<v_{F}$. In terms of normalized frequencies this condition reads 
\begin{equation}
\tilde{\omega}-\tilde{k}-\tilde{k}^{2}<0.  \label{Cond-2}
\end{equation}%
Simultaneously requiring conditions (\ref{Cond-1}) and (\ref{Cond-2}) to
hold give us the regime where two-plasmon processes are present, but the
poles of linear theory are absent. Next we would like to study three-plasmon
processes. In that case we are particularly interested in cases where
two-plasmon processes are forbidden but three-plasmon processes are allowed.
In this case we demand $v_{\text{res,-3}}<v_{F}<v_{\text{res,-2}}$ which is
written as%
\begin{equation}
\tilde{\omega}-\tilde{k}-\tilde{k}^{2}>0  \label{Cond-3}
\end{equation}%
and 
\begin{equation}
\tilde{\omega}-\tilde{k}-3\tilde{k}^{2}/2<0.  \label{Cond-4}
\end{equation}%
when $H=1$.

\begin{figure}[tbp]
\includegraphics[width=\linewidth]{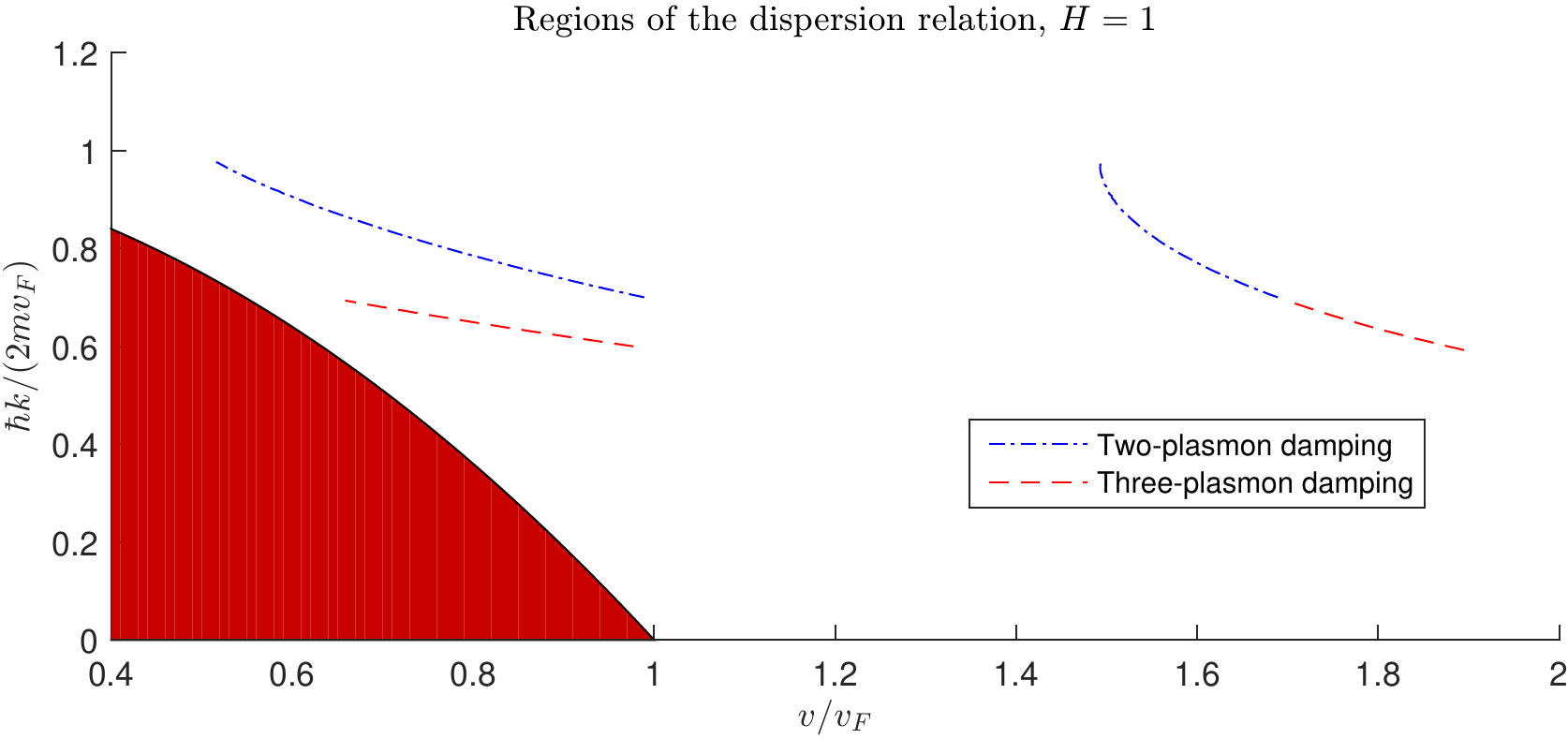}
\caption{The figure is based on the numerical solution for the dispersion
relation with $H=1$. It shows the regions of $k$-space corresponding to
two-plasmon damping (blue curve) and three-plasmon damping (red curve). In
the right part of the curves $v=\protect\omega/k$, and in the left parts the
curve has been adjusted to the resonant velocity $v=\protect\omega/k-n\hbar
k/2m$, where $n=2$ for two-plasmon damping and $n=3$ for three-plasmon
damping. For comparison the red region shows the background distribution.}
\label{fig:dispersionRegions}
\end{figure}

We are now interested in separating the dispersion curves into the
two-plasmon part (points simultaneously satisfying (\ref{Cond-1}) and (\ref%
{Cond-2})), the three-plasmon part (points simultaneously satisfying (\ref%
{Cond-3}) and (\ref{Cond-4})), and the other parts. How such a division
turns out for $H=1$ is illustrated in \cref{fig:dispersionRegions}. Picking
specific points $(\tilde{k},\tilde{\omega})$ of these curves determine the
numerical values of the coefficients in the two-plasmon and three-plasmon
system of equations that will be derived in the next section.

\subsection{The amplitude expansion}

If the nonlinearity is weak, we can use the slowly varying amplitude
approximation. In this approximation, all quantities have the form of linear
combinations of plane waves where the amplitude vary slowly as a function of
time compared to the natural frequency $\omega$. We will work to cubic order
in the amplitude as this is the lowest order that exhibits multi-plasmon
resonances. To cubic order, we can find resonances according to~%
\cref{eq:resonancecondition} with $n = 2, 3$ (again, $n = 1$ corresponds to
linear theory). In principle, we are interested in resonances with arbitrary 
$n$, but to explicitly compute them for an arbitrary $n$ means carrying out
an amplitude expansion to all orders.

To obtain evolution equations up to cubic order, we need an Ansatz for the
electrostatic potential that contains second harmonic terms and
low-frequency terms. For homogenous initial conditions, particle
conservation excludes a low-frequency contribution to the potential. Thus,
we will take the electric potential to be of the form 
\begin{equation}
\Phi = \Phi_1(t) \exp\big [ i(k z - \omega t) \big] + \Phi_2(t) \exp\big [2
i(k z - \omega t) \big]+\cc .  \label{eq:potential}
\end{equation}
Here $\cc$ denotes the complex conjugate, which is added to ensure that the
potential is real.


Similarly, the Wigner function is taken to be a periodic function 
\begin{equation}
f = F_0 + f_0(\mathbf{v}, t) + \sum_{n=1}^\infty f_n(\mathbf{v}, t) \exp [
in(k z - \omega t)] + \cc  \label{eq:f-expansion}
\end{equation}
Here $F_0$ is the initial (constant) background distribution and $f_0(%
\mathbf{v}, t)$ is the nonlinearly induced background modification. Keeping
only up to cubic terms in the amplitude expansion will allow us to truncate
the summation, but we keep it general for now to show the structure of the
equations.

We stress again that the time dependence of the amplitudes $\Phi_1, \Phi_2$
and $f_n$ is assumed to be slow compared to $\omega$.

Inserting these \emph{Ans\"{a}tze} into~\cref{eq:wigner} we will obtain a
hierarchy of equations for $\Phi_1, \Phi_2$, and the $f_n$. With a potential
of the form~\cref{eq:potential} the integral over $\mathbf{r}^{\prime }$ in~%
\cref{eq:wigner} will produce a linear combination of Dirac delta functions.
Viz., the terms in $\Phi(\mathbf{r} + \mathbf{r} ^{\prime }/2) - \Phi(%
\mathbf{r} - \mathbf{r} ^{\prime }/2)$ proportional to $e^{i(kz - \omega t)} 
$ are $\Phi_1 e^{i(kz-\omega t)}( e^{ikz^{\prime }/ 2 } - e^{-ikz^{\prime }/
2} )$ and 
\begin{align}
\int \frac{d^3{\mathbf{r} ^{\prime }} d^3 {\mathbf{v} ^{\prime }} }{%
(2\pi\hbar)^3} e^{i \mathbf{r} ^{\prime }\cdot\frac{m( \mathbf{v} - \mathbf{v%
} ^{\prime })}{\hbar} } \big[ \Phi_1 e^{i(kz-\omega t)}( e^{ikz^{\prime }/ 2
} - e^{-ikz^{\prime }/ 2} ) \big] f & = \int \frac{d^3{\mathbf{r} ^{\prime }}
d^3 {\mathbf{v} ^{\prime }} }{(2\pi\hbar)^3} \big [ e^{i \mathbf{r} ^{\prime
}\cdot \frac{m( \mathbf{v} - \mathbf{v} ^{\prime }+ \mathbf{v} _q)}{\hbar} }
-e^{i \mathbf{r} ^{\prime }\cdot\frac{m( \mathbf{v} - \mathbf{v} ^{\prime }- 
\mathbf{v} _q)}{\hbar} } \big ] \Phi_1 e^{i(kz-\omega t)} f  \notag \\
& = \int d^3 \mathbf{v} ^{\prime }\big [ \delta(\mathbf{v }- \mathbf{v}%
^{\prime }+ \mathbf{v}_q ) - \delta(\mathbf{v }- \mathbf{v}^{\prime }- 
\mathbf{v}_q) \big] \Phi_1 e^{i(kz-\omega t)} f  \notag \\
& = \Phi_1 e^{i(kz - \omega t)} \D_1 f.
\end{align}
Here we have introduced a quantum velocity shift $\mathbf{v} _q$ and the
velocity shift operator $\D_n$, defined by 
\begin{equation}
\mathbf{v} _q = \frac{\hbar k}{2m}\hat z \quad \text{and} \quad (\D_n f)(%
\mathbf{v} ) = f(\mathbf{v} + n\mathbf{v} _q) - f(\mathbf{v} - n\mathbf{v}
_q).
\end{equation}
Similarly, the terms in the potential proportional to $e^{-i(kz-\omega t)},
e^{2i(kz-\omega t)}, e^{-2i(kz-\omega t)}$ will give $-\Phi_1^* e^{-i(kz -
\omega t)} \D_1 f$, $\Phi_2 e^{2i(kz-\omega t)} \D_2 f$, and $\Phi_2^*
e^{-2i(kz-\omega t)} D_2 f$, respectively, where the star denotes complex
conjugate. 

Since the amplitudes vary on timescales much longer than $\omega$, we can
multiply any expression by $e^{-mi(kz-\omega t)}$ and average over one wave
period to pick out the terms proportional to $e^{mi(kz-\omega t)}$, where $m$
is an integer. (Essentially, we are equating Fourier components and using
the convolution theorem.) Applying this to the Wigner equation~%
\cref{eq:wigner} and using the previous results, we find the hierarchy for
the $f_n$. 
\begin{subequations}
\begin{align}
\partial_t f_0 & = \frac{iq}{\hbar}(\Phi_1 \D_1 f_1^* - \Phi_1^* \D_1 f_1 +
\Phi_2 \D_2 f_2^* - \Phi_2^* \D_2 f_2)  \label{eq:nonlinearI} \\
\partial_t {f_1} -i(\omega - kv_z) f_1 & = \frac{iq}{\hbar}( \Phi_1 \D_1(
F_0 + f_0) - \Phi_1^* \D_1 f_2 + \Phi_2 \D_2 f_1^* - \Phi_2^* \D_2 f_3 )
\label{eq:linearize} \\
\partial_t f_n - in(\omega - kv_z) f_n & = \frac{iq}{\hbar} (\Phi_1 \D_1
f_{n-1} - \Phi_1^* \D_1 f_{n+1} + \Phi_2 \D_2 f_{n-2}^* - \Phi_2^* \D_2
f_{n+2}) \quad n > 1.  \label{eq:nonlinearIII}
\end{align}
These equations can be summarized as $f_n$ coupling to $\D_1 f_{n\mp 1}$
through $\Phi_1$ ($\Phi_1^*$) and to $\D_2 f_{n \mp 2}$ through $\Phi_2$ ($%
\Phi_2^*$), if one takes $f_{-n} = f_n^*$ (as is the case for the Fourier
series of a real function).

So far, we have kept terms in the Wigner function to all orders. For $n\ge 3$%
, $f_n$ is at least cubic in the amplitude according to~%
\cref{eq:nonlinearIII}; $f_0$ and $f_2$ are quadratic. We thus only need
equations for the first three Fourier components of the Wigner function, and
can discard some terms on the right hand side that are higher order in the
amplitude expansion. Furthermore, since the problem is 1-dimensional in
velocity space, we can integrate over $v_x$ and $v_y$ and work with the
reduced Wigner function and background distribution, 
\end{subequations}
\begin{equation}
g_n(v_z) = \int dv_x dv_y \, f_n \quad \text{and} \quad G_0(v_z) = \int dv_x
dv_y \, F_0.
\end{equation}

Our final evolution equations for the Wigner function, valid to cubic order
in the amplitude, then read 
\begin{subequations}
\label{eq:g-equations}
\begin{align}
\partial_t g_0 & = \frac{iq}{\hbar}(\Phi_1 \D_1 g_1^* - \Phi_1^* \D_1 g_1)
\label{eq:new-1} \\
\partial_t {g_1} - i\delta\omega g_1 & = \frac{iq}{\hbar}( \Phi_1 \D_1( g_0
+ G_0) - \Phi_1^* \D_1 g_2+\Phi_2 \D_2 g_{1}^*)  \label{eq:new-2} \\
\partial_t g_2 - 2i \delta\omega g_2 & = \frac{iq}{\hbar} ( \Phi_1 \D_1 g_1
+\Phi_2 \D_2 G_{0}) .  \label{eq:new-3}
\end{align}
where $\delta\omega = \omega - kv_z$. We note that so far we have made no
assumptions regarding the background distribution.

\section{Multi-plasmon damping processes}

Now, the idea is to divide velocity space into a resonant region and a
non-resonant region. Unless we are close to the resonant velocity, we can
treat $\partial_t g_1$ and $\partial_t g_n$ as small corrections in~%
\cref{eq:g-equations}. The part of velocity space where this approximation
is applicable is the \emph{non-resonant region}, and the part where the time
derivative is essential for the solution is the \emph{resonant region}. The
size of the resonant region is not sharply defined. It should be taken large
enough, i.e., such that $g_n$ is negligible near the edge of the resonant
region as compared to the center, but still small compared to $v_F$, the
size of the whole velocity distribution It is possible to fulfill both
conditions, according to our numerical simulations, and our results are not
sensitive to the exact size of the resonant region.

In the non-resonant region, we solve Eqs. (\ref{eq:new-2}) and (\ref%
{eq:new-3}) for $g_1$ and $g_2$ to lowest order by neglecting the time
derivative and dividing by $-in\delta \omega $. Recursively inserting these
solutions, the repeated action of the velocity shift operators $\D_{1}$ and $%
\D_{2}$ will induce resonances for velocities fulfilling~%
\cref{eq:resonancecondition}, that is, multi-plasmon resonances.

We will treat the two-plasmon and three-plasmon cases separately, starting
with the two-plasmon case. As can be realized, when two-plasmon processes
are allowed three-plasmon processes will also take place (see Figure~\ref%
{fig:dispersionRegions}), and the damping rates are of the same magnitude.
Thus in principle two- and three-plasmon processes should be studied
simultaneously. However, the physics is easier to understand when either is
studied in isolation. In practice the wave damping due to the two processes
can be added together afterwards, which is the approach we will chose. We
note that since there is a regime that forbids two-plasmon processes but
allows three-plasmon processes, the latter can be studied in isolation, in a
cubic order amplitude expansion.

\subsection{The two-plasmon case}

We first concentrate on the two-plasmon case, where $\omega /k-v_q > v_F$
(no linear resonances) but $\omega/k-2v_{q}<v_{F}$. In this case only $g_{2} 
$ couples directly to the background distribution, and it will be resonant
when $\delta \omega \simeq 0 $ in~\cref{eq:new-3}. This means that $g_{2}$
couples resonantly to $G_{0}(\omega/k \pm 2v_{q})$ through the velocity
shift operator of the second harmonic term $\propto \Phi _{2}$. Since $%
g_{2}(v_{z})$ couples to $g_{1}(v_{z}\pm v_{q})$, $g_{1}$ will need to be
treated as resonant in two regions. Figure~\ref{fig:couplings} represents
the two-plasmon process schematically, where the arcs illustrates the
couplings induced by the velocity shift operators.


\begin{figure}[bp]
\includegraphics[width=\linewidth]{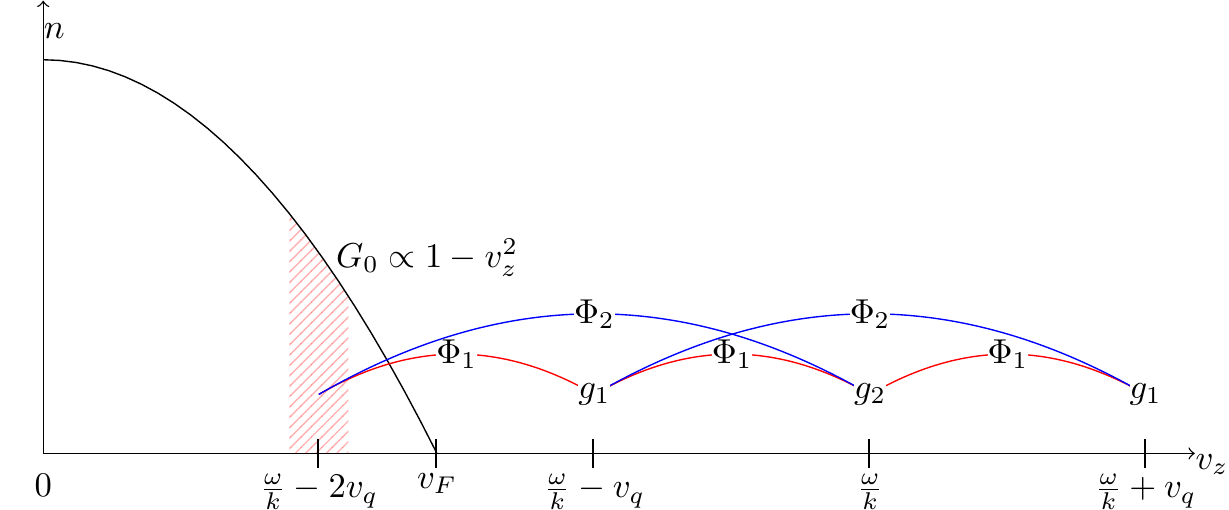}
\caption{How the components of the Wigner function couple in the 2-plasmon
damping process. The resonant region for the background distribution is
hatched, which means that $g_1$ is resonant for velocities $v_z=\protect%
\omega/k - v_q$ and $v_z=\protect\omega/k +v_q $and $g_2$ is resonant for
the velocity $v_z=\protect\omega/k$, as indicated in the figure. 	\label{fig:couplings}}
\end{figure}


To close the system, we use Poisson's equation to derive evolution equations
for the potential. Using the division into resonant and non-resonant
regions, the component proportional to $e^{2i(kz-\omega t)}$ is 
\end{subequations}
\begin{equation}
4k^2\Phi_2 = \frac{q}{\epsilon_0} \int_\text{nr} g_2 \, dv_z + \frac{q}{%
\epsilon_0} \int_\text{res} g_2 \, dv_z.
\end{equation}
In the non-resonant region, where $\delta \omega$ is large, the time
derivative on $g_2$ is negligible, and we can substitute the linear
expression for $g_1$ in the evolution equation~\cref{eq:new-3} to get 
\begin{multline}
4k^2 \Phi_2 - \frac{q^2 \Phi_2}{\epsilon_0 \hbar}\int_\text{nr} \frac{ (\D_2
G_0 )\, dv_z}{2(\omega - kv_z)} = \frac{q}{\epsilon_0} \int_\text{res} g_2
\, dv_z + \\
\frac{q^3 \Phi_1^2}{2\hbar^2\epsilon_0 } \int_\text{nr} \frac{G_0(v_z +
2v_q) - G_0(v_z)}{(\omega - kv_z)[\omega - k (v_z + v_q)]} - \frac{G_0(v_z)
- G_0(v_z - 2v_q)}{(\omega - kv_z)[\omega - k (v_z - v_q)]} \, dv_z.
\label{eq-harm}
\end{multline}
We identify the LHS of~\cref{eq-harm} as $4k^2 D(2 \omega, 2 k) \Phi_2$,
where $D(\omega, k)$ is the linear dispersion function including only the
non-resonant regions.

The coefficient for $\Phi_1^2$ can be approximated in terms of the linear
susceptibility, under the assumption that the resonant region is small. For
details, see the appendix. This approximation should be no cruder than the
ones already made, and since we have studied the system for a range of
modes, our qualitative conclusions seem robust to changes in the $\omega,k$%
-dependent coefficients. 

For convenience we introduce the notation 
\begin{equation}
\chi_n := \chi(n\omega,nk)
\end{equation}
where $\chi$ is the linear susceptibility. Note that the expression~%
\cref{suscept-1} is only an approximation when computing $\chi_2$ and $%
\chi_3 $, as these expressions have pole contributions which have been
dropped when deducing ~\cref{suscept-1}. Unless the pole contributions are
small $\chi_2$ and $\chi_3$ should be computed from the susceptibility given
in~\cref{eq:dispRel-2} rather than that given in~\cref{suscept-1}. Since $%
\omega, k$ obey the linear dispersion relation, $\chi_n = 1$, ~\cref{eq-harm}
can be written 
\begin{equation}
(1+\chi_2)\Phi_2 = -\frac{q (\chi_2 + 1/4)}{\hbar k v_q} \Phi_1^2 + \frac{q}{
\epsilon_0 k^2} \int_\text{res} g_2 \, dv_z.  \label{eq:secondHarmonic}
\end{equation}
This equation determines $\Phi_2$ in terms of the other variables.

Now we turn to the nonlinear back-reaction on $\Phi_1$. Poisson's equation
gives 
\begin{equation}
\epsilon_0 k^2 \Phi_1 = q\int g_1 \, dv_z = q\int_\text{res} g_1 \, dv_z +
q\int_\text{nr} g_1 \, dv_z.
\end{equation}
%
Here by resonant region, we mean the region of velocity space where $\delta
\omega \approx \pm kv_q$. This region is where the source term $\D_1 g_2$
has a contribution from $g_2$ evaluated near $\delta \omega = 0$.

In the non-resonant region we write 
\begin{equation}
(\omega - kv_z) g_1 = -i\frac{\partial g_1}{\partial t} - \frac{q\Phi_1}{
\hbar}\D_1 G_0 - \frac{q\Phi_2}{\hbar}\D_2 g_1^* + \frac{q \Phi_1^*}{\hbar} %
\D_1 g_2 .  \label{eq:nonResSource}
\end{equation}
Since the linear result $g_1^{\text{L}} \propto \Phi_1/(\omega - kv_z)$
holds to quadratic order, and time derivatives are assumed slow, we can use
this in the term in the small (but important) term $-i\partial_t g_1$ and
still be correct to cubic order. 
Inserting this into Poisson's equation, we obtain 
\begin{multline}
\left(1 + \frac{q^2}{\epsilon_0 \hbar k^2 } \int_\text{nr} \frac{\D_1 G_0}{%
\omega - k v_z} \, dv_z \right) \Phi_1 - i\frac{q^2}{\epsilon_0 \hbar k^2}
\int_\text{nr} \frac{\D_1 G_0}{(\omega - kv_z)^2} \, dv_z \frac{\partial
\Phi_1}{\partial t} = \\
\frac{q^2}{\epsilon_0 \hbar k^2} \int_\text{nr} \frac{1}{\omega - kv_z} (
\Phi_1^* \D_1 g_2 - \Phi_2 \D_2 g_1^* ) \, dv_z + \frac{q}{\epsilon_0 k^2}%
\int_\text{res} g_1 \, dv_z
\end{multline}
As $\omega, k$ satisfy the linear dispersion relation, the first two terms
on the left-hand side sum to zero, and the last is $i\frac{\partial D}{%
\partial \omega}$. Thus the evolution equation for $\Phi_1$ is 
\begin{equation}
i \frac{\partial D}{\partial \omega} \frac{\partial \Phi_1}{\partial t} = 
\frac {q^2}{\epsilon_0 \hbar k^2} \int_\text{nr} \frac{1}{\omega - kv_z} (
\Phi_1^* \D_1 g_2 - \Phi_2 \D_2 g_1^* ) \, dv_z + \frac {q}{\epsilon_0 k^2}%
\int_\text{res} g_1 \, dv.
\end{equation}
By substituting for $g_2, g_1^*$ according to the evolution equations and
the linear result for $g_1$, the first integral is similar to that in~%
\cref{eq-harm} and can also be approximated in terms of the linear
susceptibility. This is again detailed in an appendix. In the resonant
region, where $\delta \omega \approx kv_q$, we can solve for $g_1$ using %
\cref{eq:nonResSource} but dropping the time derivative. Since only $g_2$ is
resonant, the other contributions simply modify the coefficients of the
terms from the first integral. 

After evaluating the coefficients, we have 
\begin{equation}
i \frac{\partial D}{\partial \omega} \frac{\partial \Phi_1}{\partial t} = 
\frac{q^2}{4\hbar^2 k^2 v_q^2}(27 \chi_3 - 32\chi_2 - 6) |\Phi_1|^2 \Phi_1 + 
\frac{q}{\hbar k v_q} 8\chi_2 \Phi_1^* \Phi_2 + \frac{q^2 \Phi_1^*}{%
\epsilon_0 \hbar k^2} \int_{ \delta \omega \approx \pm v_q } \frac{ g_2(v_z
+ v_q) - g_2(v_z - v_q)}{\omega - kv_z} \, dv_z.
\end{equation}
In each component of the integration region only one of the the terms in $\D %
g_2$ contributes, viz. 
\begin{align}
\int_{ \delta \omega \approx \pm v_q } \frac{ g_2(v_z + v_q) - g_2(v_z - v_q)%
}{\omega - kv_z} \, dv_z & \approx \int_{\delta\omega \approx v_q } \frac{%
g_2(v_z + v_q)}{\omega - kv_z} \, dv_z - \int_{\delta\omega \approx -v_q } 
\frac{g_2(v_z - v_q)}{\omega - kv_z} \, dv_z \\
& = \int_\text{res} \left( \frac{1}{\omega - k(v_z - v_q)} - \frac{1}{\omega
- k(v_z + v_q)} \right) g_2(v_z) \, dv_z \\
& = \int_\text{res} \frac{ 2 v_q \, g_2}{k (v_q^2 - \tilde{v}^2) } \, d%
\tilde{v}
\end{align}
where $\tilde v = v_z - \frac{\omega}{k}$ measures velocity relative to the
resonance. This gives us the final evolution equation for $\Phi_1$, 
\begin{equation}
i\frac{\partial D }{\partial \omega} \frac{\partial \Phi_1 }{\partial t} = 
\frac{q^2 A}{\hbar^2 k^2 v_q^2} |\Phi_1|^2 \Phi_1 + \frac{q B}{\hbar kv_q}
\Phi_1^* \Phi_2 + \frac{2q^2\Phi_1^*v_q}{\hbar k^3 \epsilon_0} \int_\text{%
res } \frac{ g_2 \, d\tilde{v}}{v_q^2 - \tilde{v}^2 }.
\label{eq:evolutionPhi1}
\end{equation}
where the coefficients are 
\begin{equation}
A =\frac{1}{4} (27 \chi_3 - 32\chi_2 - 6) \quad \text{and} \quad B = 8\chi_2.
\end{equation}

The first term represents a nonlinear frequency shift. Since $\Phi_2 \sim
\Phi_1^2 + \int g_2$, the second term also contains a nonlinear frequency
shift, as well as an additional coupling to $g_2$. It is the wave-particle
interaction of the coupling to $g_2$ that is responsible for the damping;
the nonlinear frequency shift turns out to be relatively unimportant.


Next we introduce normalized dimensionless variables by $t\mapsto t\gamma $, 
$v_z\mapsto (v-\omega /k)/v_{F}$, $\Phi \mapsto q\Phi /(\hbar \gamma )$ and $%
g\mapsto q^{2}gv_{q}/(\epsilon _{0}\hbar k^{2}\gamma )$. As the evolution of
the system occurs on a time-scale much longer than the natural scales ($%
\omega ^{-1},\omega _{p}^{-1}$) we take an arbitrary value of $\gamma $,
giving an arbitrary parameter $\alpha =kv_{F}/\gamma $ in the equations. 
Then \cref{eq:secondHarmonic,eq:evolutionPhi1} take the form 
\begin{subequations}
\begin{align}
\Phi_2 & = \frac{v_F}{v_q (1 + \chi_2)} \left( \int g_2 \, dv -\frac{ \chi_2
+ 1/4}{\alpha} \Phi_1^2 \right)  \label{tp1} \\
i\frac{\partial D}{\partial \tilde{v}_\phi} \frac{\partial \Phi_1}{\partial t%
} & = \frac{A}{\alpha v_q^2} |\Phi_1|^2 \Phi_1 +\frac{B}{v_q}\Phi_2 \Phi_1^*
+ 2\Phi_1^* \int \frac{g_2 \, dv }{v_q^2 - v^2}  \label{tp2}
\end{align}
and \cref{eq:new-3} becomes 
\begin{equation}
\frac{\partial g_2}{\partial t} - 2i \alpha v g_2 = i \Phi_1^2 \frac{ G_0(v
- 2v_q) }{\alpha(v - v_q)} - i\Phi_2 G_0(v - 2v_q) + \frac{i |\Phi_1|^2 g_2 
}{\alpha (v - v_q)}  \label{tp3}
\end{equation}
where the integrals are over the resonant region. The dimensionless
background distribution is given by $G_0(v) = (3 H^2\alpha/ 32v_q^2)\operatorname{max%
}(1-v^2, 0)$.

In \cref{tp3}, we have discarded terms containing the background
distribution outside the Fermi sphere, that are non-zero for finite
temperatures. However, at finite but low temperature, just outside the Fermi
sphere $v_F$, $G_0 \approx \exp\big[(T_F/T) (1-v_z^2/v_F^2 )\big]$, and the
discarded terms are therefore exponentially small. Thus the importance of
multi-plasmon damping is not restricted to the extreme case of full
degeneracy. 

\subsection{The three-plasmon case}

Next we consider three-plasmon resonances, which become dominant for
slightly longer wavelengths, such that the lowest resonance $\omega - kv_{z}
- n v_q = 0$ occurs for $n = 3$, for some velocity $v_{z}<v_{F}$. As
indicated by the previous case of two-plasmon resonances, the nonlinear
frequency shift only plays a modest role for the damping rate, and thus
these terms are dropped henceforth. Overall the calculations use the same
methods as the two-plasmon case but in this case the important component of $%
g$ is $g_1$ near $\omega - kv_z = 0$. The process is presented schematically
in Figure~\ref{fig:couplingsThree}. 
\begin{figure}[tbp]
\includegraphics[width=\linewidth]{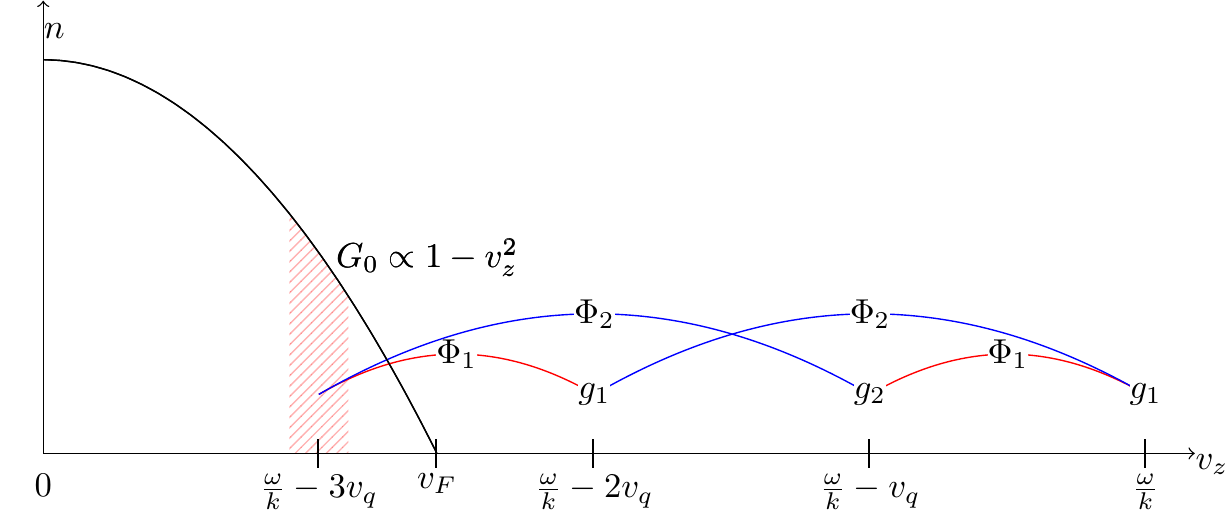}
\caption{How the components of the Wigner function couple in the 3-plasmon
damping process. The resonant region for the background distribution is
hatched, which means that $g_1$ is resonant for velocities $v_z=\protect%
\omega/k - 2v_q$ and $v_z=\protect\omega/k$ and $g_2$ is resonant for the
velocity $v_z=\protect\omega/k- v_q$, as indicated in the figure. \label{fig:couplingsThree}}
\end{figure}

With the same type of derivation as for the 2-plasmon damping, we find the
set of normalized equations 
\end{subequations}
\begin{align}
\left(1 + \chi_2 + \int_\text{res} \frac{G_0(v - 3v_q)}{\alpha(v_q - v)} \,
dv \right) \Phi_2 & = -\frac{\chi_2 + 1/4}{\alpha v_q} \Phi_1^2 + \Phi_1 
\frac{1}{\alpha} \int \frac{g_1}{v_q^2 - v^2} \, dv \\
i \frac{\partial D}{\partial v_\phi} \frac{\partial \Phi_1}{\partial t} & = 
\frac{\alpha}{v_q}\int g_1 \, dv  \label{3p1} \\
\frac{\partial g_1}{\partial t} - i \alpha vg_1 & = \frac{\chi_2 - 1/4}{
1+\chi_2}\frac{|\Phi_1|^2}{v_q^2 \alpha^2}\left (a\Phi_1 G_0(v) + bv g_1
\right)  \label{3p2}
\end{align}
where the coefficients are given by $a = \frac{3}{2}+O(v/vq)$ and $b =
1+O(v^2/v_q^2)$. We omit the higher order terms as we have already made
approximations at least this crude, and we confirm numerically that this
omission shifts the shape of $g_1$ somewhat, but does not affect the damping
rate significantly.

To obtain proper values for all parameters in the two-plasmon and
three-plasmon system respectively, we need to solve the linear dispersion
relation $\omega (k)$, which serves as the input for $\partial D/\partial
v_{\phi }$, $\chi _{2}$, etc., as shown in section II A. 

\section{Numerical results and validity}

Next we study the two-plasmon system \crefrange{tp1}{tp3} and the
three-plasmon system \crefrange{3p1}{3p2} numerically. We have implemented a
Runge-Kutta scheme~\cite{kassam2005fourth}, fourth order in time, intended
for such problems and obtained results stable against changing both the
width of the resonant region and the resolution.

\begin{figure}[tbp]
\includegraphics[width=\linewidth]{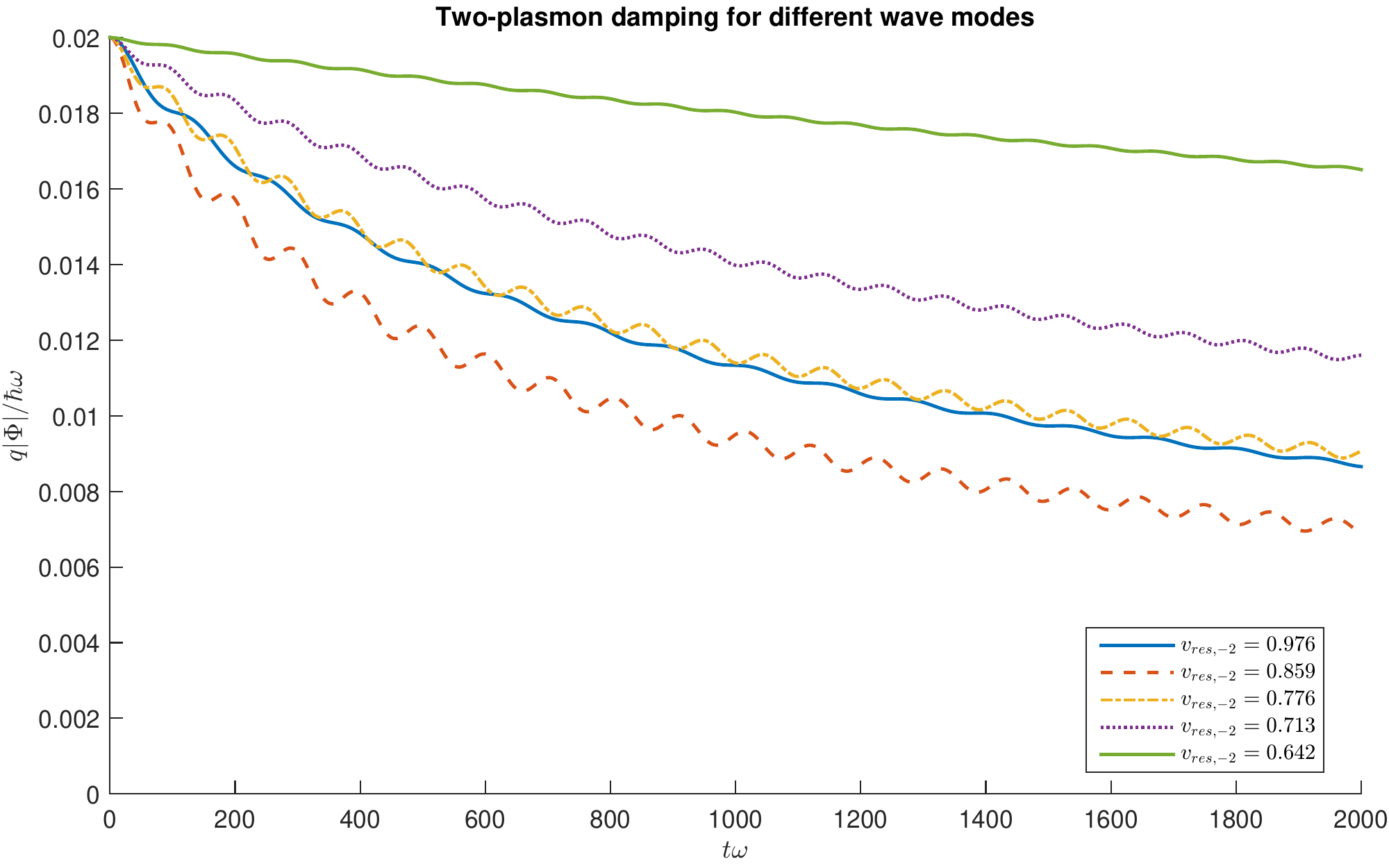}
\caption{Two-plasmon damping for different wave modes. Each mode decays
roughly as $|\Phi(t)| = |\Phi(0)|/(1+t/t_0)^{1/2}$, for different values of $%
t_0$. The small scale oscillations of each curve are not a physical effect,
but rather artifacts of the resonant region approximation. The resonant
velocities $v_{\mathrm{res},-2}$ for the different curves are normalized
against the Fermi velocity.}
\label{fig:twoPlasmonProfiles}
\end{figure}

\begin{figure}[tbp]
\includegraphics[width=\linewidth]{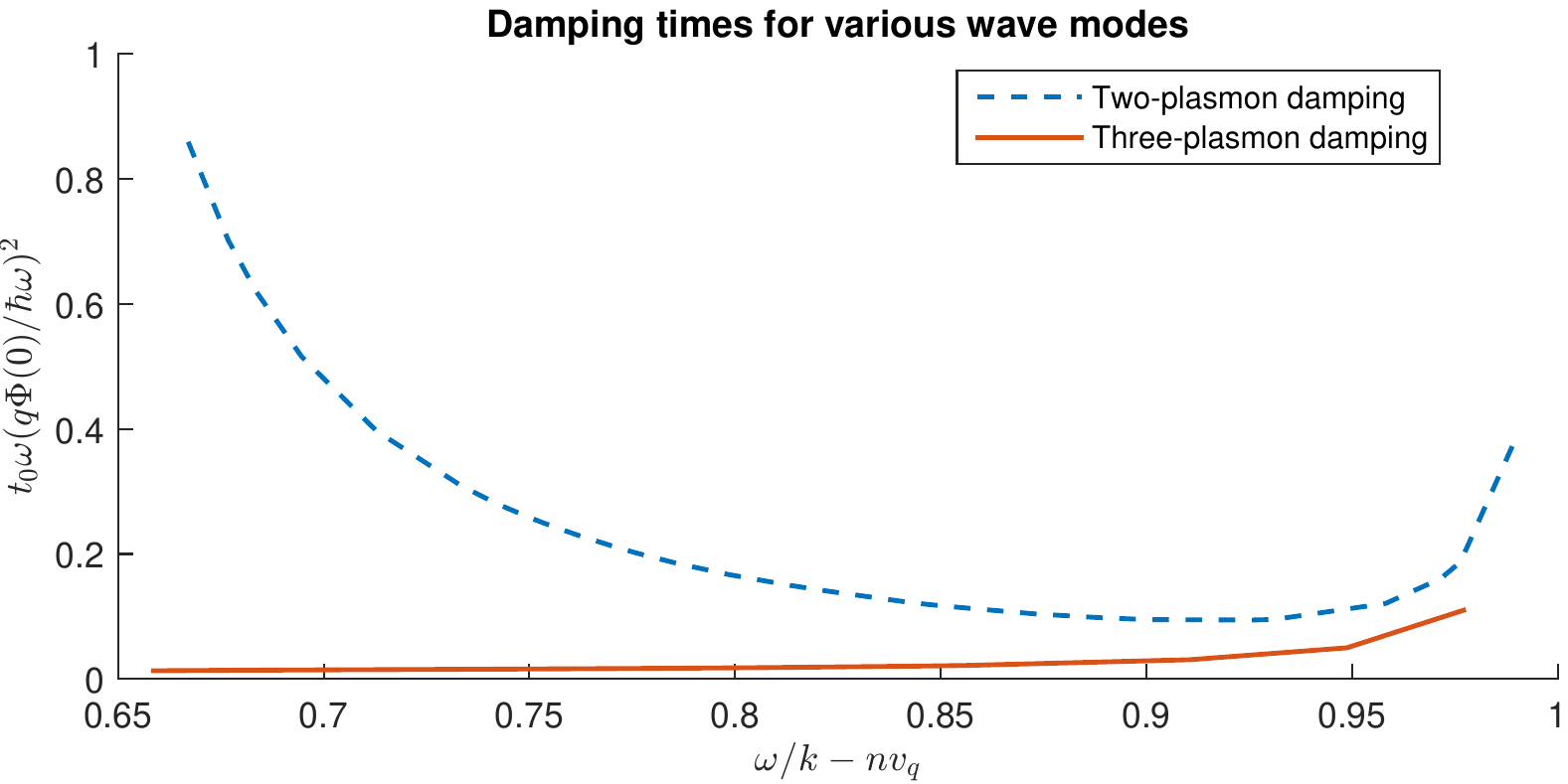}
\caption{Damping times for various wave modes for both processes considered,
as a function of the velocity of the resonant particles. The three-plasmon
curve does not extend over the same region as the two-plasmon curve because
of the shape of the dispersion relation. See section II A for details.}
\label{fig:dampingTimes}
\end{figure}

The damping of the wave amplitude scales the same in the two-plasmon and the
three-plasmon case, and in both cases the following analytical fit 
\begin{equation}
|\Phi (t)|=|\Phi (0)|/(1+t/t_{0})^{1/2}
\end{equation}%
is a good approximation. Here $t_{0}$ is a characteristic damping time that
scales as 
\begin{equation}
t_{0}=C(v_{q},\omega /k)\left\vert \frac{\hbar \omega }{q\Phi (0)}%
\right\vert ^{2}\frac{1}{\omega }.  \label{eq:dampingTime}
\end{equation}%
The amplitude evolution for some typical cases are shown in Figure~\ref%
{fig:twoPlasmonProfiles}. 
The factor $C$ varies between $0.03$ and $0.5$, depending on the precise
location of the resonance, as shown in Figure~\ref{fig:dampingTimes}. The
damping time does not decrease monotonously as the resonance is moved
further into the bulk of the background distribution as one would perhaps
expect \emph{prima facie}; this is explained by the dependence of the
coefficients in \cref{tp1}--\cref{tp3} on $k$.

\cref{fig:twoPlasmonProfiles} also shows amplitude oscillations. These are
due to a frequency shift that is an artefact of taking a finite resonant
region. Taking a larger resonant region decreases the oscillations, but if
the resonant region has to be made very large, the approximations we have
used cannot be justified anymore. For the three-plasmon case, these
artefacts are much smaller.

\begin{figure}[h]
\subfloat[Two-plasmon damping]{	\includegraphics[width=.95\linewidth]{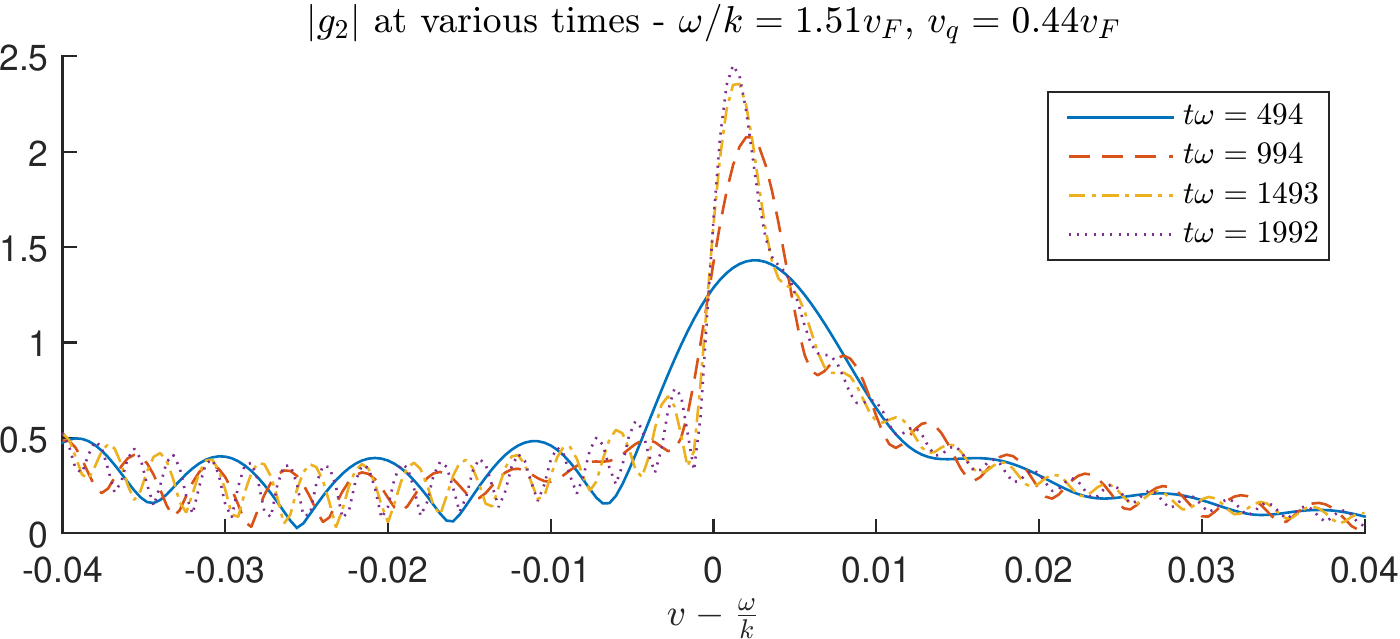}
	}
\par
\subfloat[Three-plasmon damping]{	\includegraphics[width=.95\linewidth]{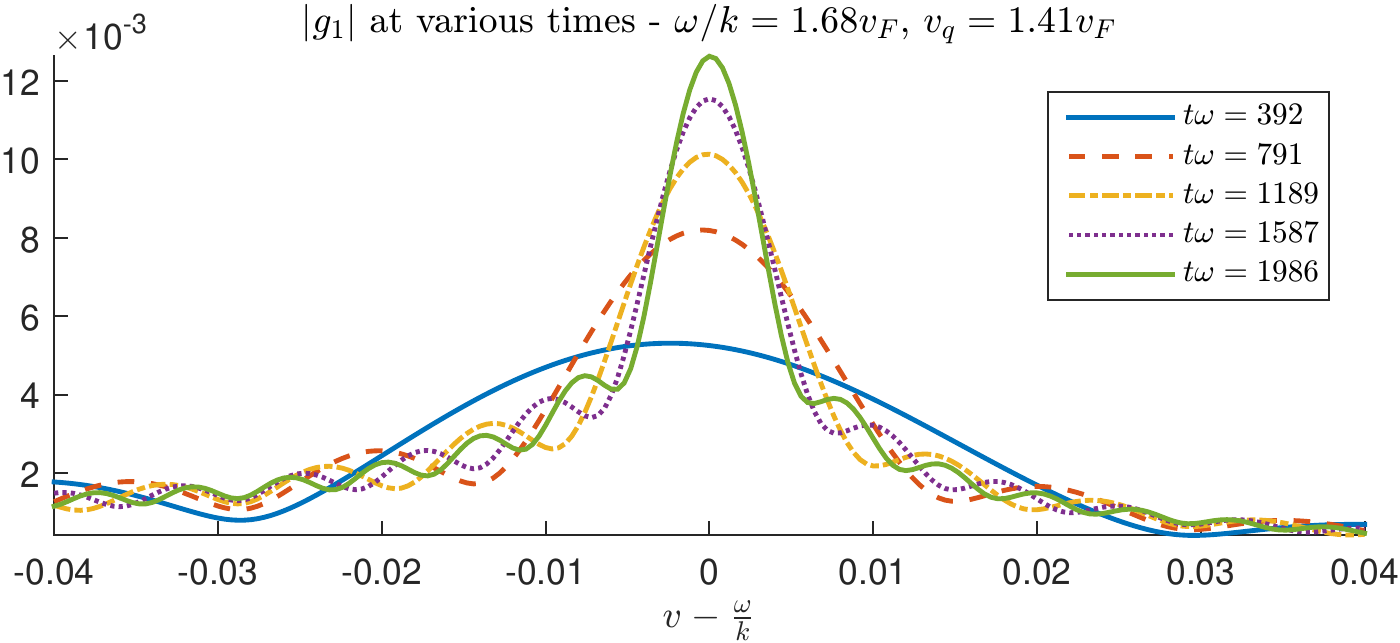}
	}
\caption{The absolute value of the resonant Fourier component of the Wigner
function at various times. It is seen that the Wigner function is
significant mainly near the resonance. At late times fine structures develop
in velocity space. }
\label{fig:wignerFunction}
\end{figure}

The term $\propto \Phi _{1}^{2}$ in \cref{tp1} gives a term $\propto \Phi
_{1}\left\vert \Phi_{1}\right\vert ^{2}$ when the expression for $\Phi_{2}$
is substituted into \cref{tp2}), i.e., a nonlinear frequency shift term.
However, this term is relatively unimportant and solving \crefrange{tp1}{tp3}
with or without this term, does not alter the damping rate significantly.
Thus the omission of the nonlinear frequency shift term when studying
three-plasmon processes is justified -- at least as far as we are only
concerned with the evolution of the wave amplitude. If we shift our interest
to the evolution of the resonant particles, see \cref{fig:wignerFunction},
the situation is somewhat changed. Here a frequency shift changes the exact
location of the resonance. Since the approximation of a finite resonant
region also gives rise to a frequency shift, this is still seen in %
\cref{fig:wignerFunction}. This effect is much less pronounced in the
three-plasmon case, as the size of the resonant region is less relevant in
this case. 

\section{Conclusion and discussion}

In the present paper we have studied wave-particle interaction due to
multi-plasmon resonances. We note that the appearance of these new
resonances is independent of the background distribution, and hence our
results can be immediately generalized to cases with a finite temperature.
At finite temperature, there are, however, always linearly resonant
particles in the tail of the Fermi-Dirac distribution. Since the tail
outside the Fermi sphere is exponentially small in $T/T_F$ we can expect the
physics to be the same for low temperatures as for zero temperatures. In
effect, by taking a completely degenerate plasma we have only ignored
exponentially few particles, compared to the many particles near the
multi-plasmon resonances.

A more quantitative estimate comes from comparing the linear dispersion
relation based on Eq. (\ref{suscept-1}) with that computed for an arbitrary
degeneracy, see Ref.~\cite{Rightley-Uzdensky}. For low temperatures, $T
\lesssim 0.2T_F$, the linear damping rate $\gamma$ is of the order $10^{-3}
\omega_p$ or even smaller for the range of wave numbers considered in this
paper. In \cref{fig:twoPlasmonProfiles}, we can see that multi-plasmon
damping happens on a similar timescale. For higher initial amplitudes
nonlinear multi-plasmon damping is faster, while the linear damping time is
the same, as seen from Eq. \cref{eq:dampingTime}. This shifts the relative
importance toward multi-plasmon damping when the amplitude is larger the
allowed limit, resulting from the assumption that the resonant region is
small.

The separation of the resonant velocity between the linear resonance and the
multi-plasmon resonances with $n=2 , 3, \ldots $ is directly proportional to the
standard quantum parameter $H= \hbar\omega_p/E_k$. Here $E_k$ is the
characteristic kinetic energy of the particles, given by $E_k=k_B T_F$ for a
degenerate plasma and $E_k=k_B T$ in the non-degenerate case. For plasma
densities of the order of metallic systems, $H$ is of the order of unity
when $T\ll T_F$, and hence the resonances for e.g. $n=1$ and $n=3$ are well
separated. Increasing the temperature, $H$ is kept of the order unity at
least up to $T\sim T_F$. However, increasing the temperature further to
temperatures $T\gg T_F$, the relative importance of the new resonances
gradually diminishes, since the separation from the classical resonances
become small. 

For $T\ll T_{F}$, two plasmon processes occur for wavenumbers $k_{2}<k<k_{%
\text{cr}}\ $, where $k_{2}\simeq 0.69\omega _{p}/v_{F}$ and $k_{\text{cr}%
}=0.98\omega _{p}/v_{F}$. Whenever two-plasmon processes occur, we also have
three-plasmon processes, which give a damping of comparable magnitude. A
more precise comparison of two-and three-plasmon damping is made in %
\cref{fig:dampingTimes}. For $k_{3}<k<k_{2}$, where $k_{3}=0.60\omega
_{p}/v_{F}$, three-plasmon damping dominates and two-plasmon processes are
absent. 
This holds because $n$-plasmon processes with $n\geq 4$ are of higher order
in an amplitude expansion. For multi-plasmon processes with $n=2$ or $n=3$,
the damping rate decreases with the amplitude, as given by %
\cref{eq:dampingTime}. Thus these processes are of limited importance for
very small amplitudes, in which case collisional damping will be the
dominant damping mechanism when $k<k_{\text{cr}}$.

In the present paper we have been concerned with damping mechanisms, but
just like standard Landau damping can turn into an instability by modifying
the initial distribution, the same is true for multi-plasmon resonances. In
this context it should be noted that generally for resonant velocities in Eq.\eqref{eq:resonancecondition}, the negative signs in the expression are
associated with resonant particles absorbing wave quanta from the wave
(leading to wave damping), whereas the positive signs are associated with
particles emitting quanta to the wave (leading to wave growth), if the wave is propagating in the positive direction.
In our calculation we have started from a background distribution in thermodynamic equilibrium (more specifically a fully degenerate Fermi-Dirac distribution), and have chosen a wavenumber with resonant velocities such that no resonant particles with the positive sign in Eq.~\eqref{eq:resonancecondition} are present initially.
Actually, during the wave damping process particles with
a velocity equal to $v_{\mathrm{res},-2}$ (for two-plasmon
damping) or $v_{\mathrm{res},-3}$ (for three-plasmon damping) are
transferred to velocities $v_{\mathrm{res},2}$ and $v_{\mathrm{res},3}$
respectively, when absorbing multiple wave quanta. However, for the modest
initial amplitude considered in our paper this is a small effect affecting
only a limited number of particles, that does not deplete the background
distribution significantly. As a consequence the nonlinear modification of
the background distribution that takes place is not sufficient to reverse
the damping process. However, for a different background distribution where
the initial number of particles is larger at the positive resonant velocity
(i.e. a background distribution where $f_{0}(v_{\text{res},+n})>f_{0}(v_{%
\text{res},-n})$) we should expect the process to run in the opposite
direction, i.e. we should have a wave instability rather than wave damping.
In the limit where $\hbar k/2m \to 0$ this turns into the standard
bump-on tail instability that follows from a negative slope of the
distribution function. However, we will not pursue this issue here and a
study of instabilities associated with multi-plasmon resonances is a project
for further research.

In conclusion, the processes studied here can be important for broad classes
of systems. In particular multi-plasmon resonances can play a role in dense
plasmas such as metallic plasmas, white dwarf stars and inertially confined
plasmas. More generally, the methods used here can be used to study resonant
interaction involving multiple wave quanta of other types of waves, which
can be of significance for wave-particle damping of any wave mode, provided
the wavelength is not much longer than the de Broglie wavelength. The
generality of the mechanism studied here follows from the fact that~\cref{eq:resonancecondition} can be derived from energy-momentum
conservation only. As a consequence, resonant wave-particle interaction in
high-density plasmas involving intense short-wavelength electromagnetic
radiation can be possible for $n\geq 1$, potentially leading to new
opportunities for particle acceleration. However, this remains a project for
further study.

\begin{acknowledgments}
	The authors would like to acknowledge financial support by the Swedish Research Council, grant number 2012-3320. G.B. and J.Z. also acknowledge financial support from Wallenberg Foundation within the grant "Plasma based compact ion sources" (PLIONA). 
\end{acknowledgments}

\appendix

\section{Detailed calculation of coefficients}

We detail how to relate the coefficient for $\Phi_1^2$ in~\eqref{eq-harm} to
the dispersion function. The coefficient is 
\begin{equation*}
\frac{ q^3 }{8\hbar^2 k^2 \epsilon_0} \int_\text{nr} \frac{G_0(v_z + 2v_q) -
G_0(v_z)}{(\omega - kv_z)[\omega - k (v_z + v_q)]} - \frac{G_0(v_z) -
G_0(v_z - 2v_q)}{(\omega - kv_z)[\omega - k (v_z - v_q)]} \, dv_z.
\end{equation*}
which by partial fractions decomposition is a sum of terms of the form $\int_%
\text{nr} G_0(v_z)/(\omega - kv_z + mv_q)$ where $m$ is an integer. These
integrands are of the same form as in the linear dispersion function~%
\eqref{eq:dispRel-1}, but the integral is over the non-resonant region only.
Since by assumption the resonant region is small, it is valid to approximate 
$\int_\text{nr} \approx \int_\text{nr} + \int_\text{res}$ and use the
expression~\eqref{suscept-1}, in case the pole contribution is small.
Otherwise the susceptibility should be computed from \eqref{eq:dispRel-3}.

Concretely, the two terms in the integrand as 
\begin{align}
\frac{G_0(v_z + 2v_q) - G_0(v_z)}{(\omega - kv_z)[\omega - k (v_z + v_q)]} &
= \frac{G_0(v_z+2v_q) - G_0(v_z)}{kv_q} \left( \frac{1}{\omega - k(v_z +
v_q) } - \frac{1}{\omega - kv_z} \right) \\
\frac{G_0(v_z) - G_0(v_z - 2v_q)}{(\omega - kv_z)[\omega - k (v_z - v_q)]} &
= \frac{G_0(v_z) - G_0(v_z - 2v_q)}{kv_q} \left(\frac{1}{\omega - kv_z} - 
\frac{1}{\omega - k(v_z - v_q) } \right)
\end{align}
Now, by shifting the integration variable it is seen that 
\begin{equation}
\frac{ q^3 }{8\hbar^2 k^2 \epsilon_0 v_q} \int_\text{nr} \frac{G_0(v_z \pm
2v_q) - G_0(v_z)}{\omega - k(v_z \pm v_q)} \, dv_z = \frac{q}{8\hbar k v_q }
\chi_1 = -\frac{q}{8\hbar k v_q }
\end{equation}
where the last step is that $\chi_1 = -1$, as $\omega, k$ verify the linear
dispersion relation. Furthermore, the $G_0(v_z)/(\omega - kv_z)$ terms enter
with opposite signs and cancel. Therefore, the coefficient is 
\begin{equation}
-\frac{q}{4\hbar k v_q } - \frac{ q^3 }{8\hbar^2 k^3 \epsilon_0 v_q}\int_%
\text{nr} \frac{G_0(v+2v_q) - G_0(v - 2v_q)}{\omega - kv_z} = -\frac{q}{%
4\hbar k v_q } - \frac{q}{\hbar kv_q} \chi_2 = -\frac{q}{\hbar kv_q} \bigg ( 
\frac{1}{4} + \chi_2\bigg).
\end{equation}

We have 
\begin{multline}
\frac{q^2\Phi_1^*}{\epsilon_0 \hbar k^2} \D_1 g_2 = -\frac{q^3\Phi_1^*}{%
\epsilon_0 \hbar^2 k^2} \bigg[ \frac{g_1(v_z + 2v_q) -g_1(v) }{\omega - k
(v_z + v_q)}\Phi_1 + [G_0(v_z + 3v_q) - G_0(v-v_z) ] \Phi_2 \\
- \frac{g_1(v) - g_1(v_z - 2v_q) }{\omega - k (v_z - v_q)}\Phi_1 -
[G_0(v+v_z) - G_0(v - 3v_z)] \Phi_2 \bigg]
\end{multline}
Thus we need to evaluate for the $\Phi_1$ part, 
\begin{multline}
\frac{q^4 |\Phi_1|^2 \Phi_1}{2 \epsilon_0 \hbar^3 k^2}\int_\text{nr} \frac{1%
}{(\omega - kv_z)} \bigg[ \frac{1}{\omega - k (v_z + v_q)} \left( \frac{%
G_0(v_z + 3v_q) - G_0(v_z + v_q) }{\omega - k (v_z + 2v_q)} - \frac{G_0(v_z
+ v_q) - G_0(v_z - v_q)}{\omega - kv_z} \right) \\
- \frac{1}{\omega - k(v_z - v_q)} \left( \frac{G_0(v_z + v_q) - G(v_z - v_q)%
}{\omega - kv_z} - \frac{G_0(v_z - v_q) - G_0(v_z - 3v_q)}{\omega - k (v_z -
2v_q)} \right) \bigg] \, dv_z.
\end{multline}
Disregarding the prefactor, after a shift of integration variable and
partial fractions, the first fraction is 
\begin{equation}
\frac{1}{k^2 v_q^2}\int \left( \frac{1}{2( \omega - k(v_z - 2v_q) )} - \frac{%
1}{\omega - k(v_z - v_q)} + \frac{1}{2(\omega - kv_z)} \right) \big [ %
G_0(v_z + v_q) - G_0(v_z - v_q) \big] \, dv_z
\end{equation}
while the last fraction is 
\begin{equation}
\frac{1}{k^2 v_q^2}\int \left( \frac{1}{2( \omega - k(v_z + 2v_q) )} - \frac{%
1}{\omega - k(v_z + v_q)} + \frac{1}{2(\omega - kv_z)} \right) \big [ %
G_0(v_z + v_q) - G_0(v_z - v_q) \big] \, dv_z.
\end{equation}
These enter with the same sign. With suitable shifts of integration
variable, we can write these as 
\begin{align}
\frac{1}{k^2 v_q^2} \int \frac{G_0(v_z + 3v_q) - G_0(v_z - v_q)}{2(\omega -
kv_z)} - \frac{G_0(v_z + 2v_q) - G_0(v_z)}{\omega - kv_z} + \frac{G_0(v +
v_q) - G_0(v_z - v_q) }{2(\omega - kv_z)} \, dv_z \\
\frac{1}{k^2 v_q^2} \int \frac{G_0(v_z - v_q) - G_0(v_z - 3v_q)}{2(\omega -
kv_z)} - \frac{G_0(v_z) - G_0(v_z - 2v_q)}{\omega - kv_z} + \frac{G_0(v +
v_q) - G_0(v_z - v_q) }{2(\omega - kv_z)} \, dv_z
\end{align}
respectively. Adding these together and taking into account the prefactor
again yields 
\begin{equation}
\frac{q^2 |\Phi_1| \Phi_1 }{2\hbar^2 k^2 v_q^2} \left (\frac{27}{2} \chi_3 -
8\chi_2 - 1 \right).
\end{equation}

The middle two fractions are 
\begin{multline}
- \int_\text{nr} \frac{1}{(\omega - kv_z)^2}\left( \frac{1}{ \omega - k(v_z
+ v_q)} + \frac{1}{\omega - k(v_z - v_q)} \right) \big[G_0(v_z + v_q) -
G_0(v_z - v_q)\big] \, dv_z \\
= - \frac{1}{k^2v_q^2}\int_\text{nr} \left( \frac{1}{\omega - k(v_z+v_q)} + 
\frac{1}{\omega - k(v_z - v_q)} - \frac{2}{\omega - kv_z} \right) \big[ %
G_0(v_z + v_q) - G_0(v_z - v_q)\big] \, dv_z.
\end{multline}
We recognize the last term as giving $\chi_1$. For the first two, we shift
integration variables to yield 
\begin{multline}
-\frac{2q^2 |\Phi|^2 \Phi}{2\hbar^2 k^2 v_q^2} + \frac{q^4 |\Phi|^2 \Phi}{%
2\epsilon_0 \hbar^3 k^4 v_q^2 } \int_\text{nr} \frac{G_0(v_z- 2v_q) -
G_0(v_z) - G_0(v_z + 2v_q) + G_0(v_z) }{\omega - kv_z} \, dv_z = -\frac{q^2
|\Phi|^2 \Phi}{2 \hbar^2 k^2 v_q^2 }(2 + 8 \chi_2)
\end{multline}
This gives the total coefficient for $|\Phi_1|^2 \Phi$ as 
\begin{equation}
\frac{q^2 }{2 \hbar^2 k^2 v_q^2 } (\frac{27}{2} \chi_3 - 16 \chi_2 -3 )
\end{equation}

For $\Phi_2$, we have 
\begin{equation}
- \frac{q^3 \Phi_1^* \Phi_2}{2 \epsilon_0 \hbar^2 k^2}\int_\text{nr} \frac{%
G_0(v_z + 3v_q) - G_0(v-v_z) - G_0(v+v_z) + G_0(v - 3v_z) }{(\omega - kv_z)}
\Phi_2 \, dv_z
\end{equation}
With a partial fractions decomposition, this is 
\begin{multline}
- \frac{q^3 \Phi_1^* \Phi_2}{2\epsilon_0 \hbar^2 k^3 v_q}\int_\text{nr}
[G_0(v_z + 3v_q) - G_0(v_z - v_q) ] \bigg(\frac{1}{\omega - k(v_z + v_q)} - 
\frac{1}{\omega - kv_z} \bigg) \\
- [G_0(v_z + v_q) - G_0(v_z - 3v_q) ] \bigg(\frac{1}{\omega - kv_z} - \frac{1%
}{\omega - k(v_z - v_q)}\bigg) \\
= - \frac{q^3 \Phi_1^* \Phi_2}{2\epsilon_0 \hbar^2 k^3 v_q}\int_\text{nr} 
\frac{G(v_z + 3v_q) - G(v_z - 3v_q)}{\omega - kv_z} + \frac{G_0(v_z + 3v_q)
- G_0(v_z - v_q)}{\omega -k(v_z + v_q)} \\
+ \frac{G_0(v_z + v_q) - G_0(v_z - 3v_q)}{\omega -k(v_z - v_q)} + \frac{%
G_0(v_z - v_q) - G_0(v_z + v_q)}{\omega - kv_z}\, dv_z \\
= -\frac{q \Phi_1^* \Phi_2}{2\hbar k v_q} \big( 27 \chi_3 + 8 \chi_2 + 8
\chi_2 - \chi_1 \big).
\end{multline}

Lastly, we have 
\begin{equation}
\frac{q\Phi_2}{\hbar}\D_2 g_1^* = - \frac{q^2\Phi_2\Phi_1^*}{\hbar} \frac{1}{%
\omega - kv_z}\left[ \frac{G_0(v_z + 3v_q) - G_0(v_z + v_q)}{\omega - k(v_z
+ 2v_q) } - \frac{G_0(v_z - v_q) - G_0(v_z - 3v_q)}{\omega - k(v_z - 2v_q) }%
. \right]
\end{equation}
With partial fractions, the two terms are 
\begin{align}
\frac{G_0(v_z + 3v_q) - G_0(v_z + v_q)}{(\omega - kv_z)[\omega - k(v_z +
2v_q) ]} & = \frac{G_0(v_z + 3v_q) - G_0(v_z + v_q)}{2kv_q} \left[ \frac{1}{%
\omega - k(v_z + 2v_q)} - \frac{1}{\omega - kv_z} \right] \\
\frac{G_0(v_z - v_q) - G_0(v_z - 3v_q)}{(\omega - kv_z)[\omega - k(v_z -
2v_q)] } & = \frac{G_0(v_z - v_q) - G_0(v_z - 3v_q)}{2kv_q} \left[ \frac{1}{%
\omega - kv_z} - \frac{1}{\omega - k(v_z - 2v_q)} \right]
\end{align}
Therefore, the integral is 
\begin{multline}
- \frac{q^3\Phi_2\Phi_1^*}{2\epsilon_0 \hbar^2 k^3 v_q} \int_\text{nr} \frac{%
G_0(v_z + 3v_q) - G_0(v_z + v_q)}{\omega - k(v_z + 2v_q)} - \frac{G_0(v_z +
3v_q) - G_0(v_z - 3v_q)}{\omega - kv_z} \\
+ \frac{G_0(v_z - v_q) - G_0(v_z - 3v_q)}{\omega - k(v_z - 2v_q)} + \frac{%
G_0(v_z + v_q) - G_0(v_z - v_q)}{\omega - kv_z} \, dv_z \\
= - \frac{q \Phi_2\Phi_1^*}{2\hbar kv_q} \big( \chi_1 - 27 \chi_3 + \chi_1 +
\chi_1 \big).
\end{multline}

Our conclusion is that 
\begin{multline}
\frac{q^2}{\epsilon_0 \hbar k^2} \int_\text{nr} \frac{1}{\omega - kv_z} (
\Phi_1^* \D_1 g_2 - \Phi_2 \D_2 g_1^* ) \, dv_z = \\
\frac{q^2}{4\hbar^2 k^2 v_q^2}(27 \chi_3 - 32\chi_2 - 6) |\Phi_1|^2 \Phi_1 + 
\frac{q}{2 \hbar k v_q}(27 \chi_3 + 16 \chi_2 + 1) \Phi_1^* \Phi_2 - \frac{q%
}{2 \hbar kv_q } (27 \chi_3 + 1)\Phi_1^* \Phi_2 \\
= \frac{q^2}{4\hbar^2 k^2 v_q^2}(27 \chi_3 - 32\chi_2 - 6) |\Phi_1|^2 \Phi_1
+ \frac{q}{\hbar k v_q} 8\chi_2 \Phi_1^* \Phi_2 .
\end{multline}


\end{document}